\documentclass[letterpaper]{article} 
\usepackage{aaai2026}  
\usepackage{times}  
\usepackage{helvet}  
\usepackage{courier}  
\usepackage[hyphens]{url}  
\usepackage{graphicx} 
\usepackage{natbib}  
\usepackage{caption} 
\usepackage{algorithm}
\usepackage{algorithmic}
\usepackage{booktabs}

\usepackage{newfloat}

\usepackage{listings}
\usepackage{bera}

\usepackage{xcolor}

\colorlet{punct}{red!60!black}
\definecolor{background}{HTML}{EEEEEE}
\definecolor{delim}{RGB}{20,105,176}
\colorlet{numb}{magenta!60!black}

\DeclareCaptionStyle{ruled}{labelfont=normalfont,labelsep=colon,strut=off} 
\floatstyle{ruled}
\newfloat{listing}{tb}{lst}{}
\floatname{listing}{Listing}
\title{VERAFI: Verified Agentic Financial Intelligence through Neurosymbolic Policy Generation}

\author{
    Adewale Akinfaderin\textsuperscript{\rm 1},
    Shreyas Subramanian\textsuperscript{\rm 1}
}
\affiliations{
    \textsuperscript{\rm 1}Amazon Web Services\\
    Seattle, WA, USA\\
    akinfaa@amazon.com, subshrey@amazon.com
}

\lstdefinelanguage{json}{
    basicstyle=\normalfont\ttfamily,
    numbers=left,
    numberstyle=\scriptsize,
    stepnumber=1,
    numbersep=8pt,
    showstringspaces=false,
    breaklines=true,
    frame=lines,
    backgroundcolor=\color{background},
    literate=
     *{0}{{{\color{numb}0}}}{1}
      {1}{{{\color{numb}1}}}{1}
      {2}{{{\color{numb}2}}}{1}
      {3}{{{\color{numb}3}}}{1}
      {4}{{{\color{numb}4}}}{1}
      {5}{{{\color{numb}5}}}{1}
      {6}{{{\color{numb}6}}}{1}
      {7}{{{\color{numb}7}}}{1}
      {8}{{{\color{numb}8}}}{1}
      {9}{{{\color{numb}9}}}{1}
      {:}{{{\color{punct}{:}}}}{1}
      {,}{{{\color{punct}{,}}}}{1}
      {\{}{{{\color{delim}{\{}}}}{1}
      {\}}{{{\color{delim}{\}}}}}{1}
      {[}{{{\color{delim}{[}}}}{1}
      {]}{{{\color{delim}{]}}}}{1},
}

\begin{document}

\maketitle

\begin{abstract}
Financial AI systems suffer from a critical blind spot: while Retrieval-Augmented Generation (RAG) excels at finding relevant documents, language models still generate calculation errors and regulatory violations during reasoning, even with perfect retrieval. This paper introduces VERAFI (Verified Agentic Financial Intelligence), an agentic framework with neurosymbolic policy generation for verified financial intelligence. VERAFI combines state-of-the-art dense retrieval and cross-encoder reranking with financial tool-enabled agents and automated reasoning policies covering GAAP compliance, SEC requirements, and mathematical validation. Our comprehensive evaluation on FinanceBench demonstrates remarkable improvements: while traditional dense retrieval with reranking achieves only 52.4\% factual correctness, VERAFI's integrated approach reaches 94.7\%, an 81\% relative improvement. The neurosymbolic policy layer alone contributes a 4.3 percentage point gain over pure agentic processing, specifically targeting persistent mathematical and logical errors. By integrating financial domain expertise directly into the reasoning process, VERAFI offers a practical pathway toward trustworthy financial AI that meets the stringent accuracy demands of regulatory compliance, investment decisions, and risk management.
\end{abstract}


\section{Introduction}

Financial artificial intelligence systems operate in high-stakes environments where accuracy is paramount for regulatory compliance, investment decisions, and risk management. While Retrieval-Augmented Generation (RAG) has emerged as a leading approach for grounding large language models in external knowledge \citep{lewis2020retrieval}, financial applications present unique challenges that extend beyond traditional knowledge-intensive tasks. Financial documents contain complex numerical relationships, temporal dependencies, and regulatory constraints that demand not only accurate retrieval but also mathematically precise reasoning and compliance validation. The exponential growth in financial data complexity, combined with stringent accuracy requirements for applications such as SEC filing analysis, GAAP compliance checking, and investment advisory services, necessitates AI systems that can provide enhanced accuracy rather than probabilistic confidence in their outputs.

Recent advances in financial RAG have demonstrated significant improvements through enhanced retrieval strategies and agentic AI integration. \citet{srinivasan2025enhancing} introduced Multi-HyDE and agentic frameworks specifically designed for financial question-answering, achieving notable accuracy improvements and hallucination reduction on financial benchmarks through query decomposition and tool orchestration. However, even these enhanced approaches face a fundamental limitation: they cannot prevent post-generation errors that occur during the reasoning phase, even when correct financial documents are successfully retrieved. Our analysis reveals that while state-of-the-art retrieval methods using advanced embeddings and cross-encoder reranking can successfully identify relevant financial documents, and agentic systems with financial tool-use show promising performance, mathematical calculation errors, temporal inconsistencies, and regulatory compliance violations persist in the generated responses. These post-retrieval hallucinations represent a critical gap in financial AI reliability, as they occur precisely when the system has access to correct information but fails during the generation and reasoning process.

To address these challenges in financial AI reliability, we present VERAFI (Verified Agentic Financial Intelligence), a novel neurosymbolic framework that integrates state-of-the-art retrieval components with agentic tool-use and policy-guided generation to ensure mathematical accuracy and regulatory compliance. Our system combines Qwen3-Embedding-4B embeddings \citep{zhang2025qwen3}, Jina-reranker-v3 cross-encoder reranking \citep{wang2025jina}, agentic financial tools including calculators and Python execution environments, and a neurosymbolic validation layer that provides enhanced accuracy through automated reasoning policies covering GAAP compliance, SEC regulatory requirements, and mathematical accuracy validation. The main contributions of this work are:

\begin{figure*}[t]
\centering
\fbox{\includegraphics[width=0.95\textwidth]{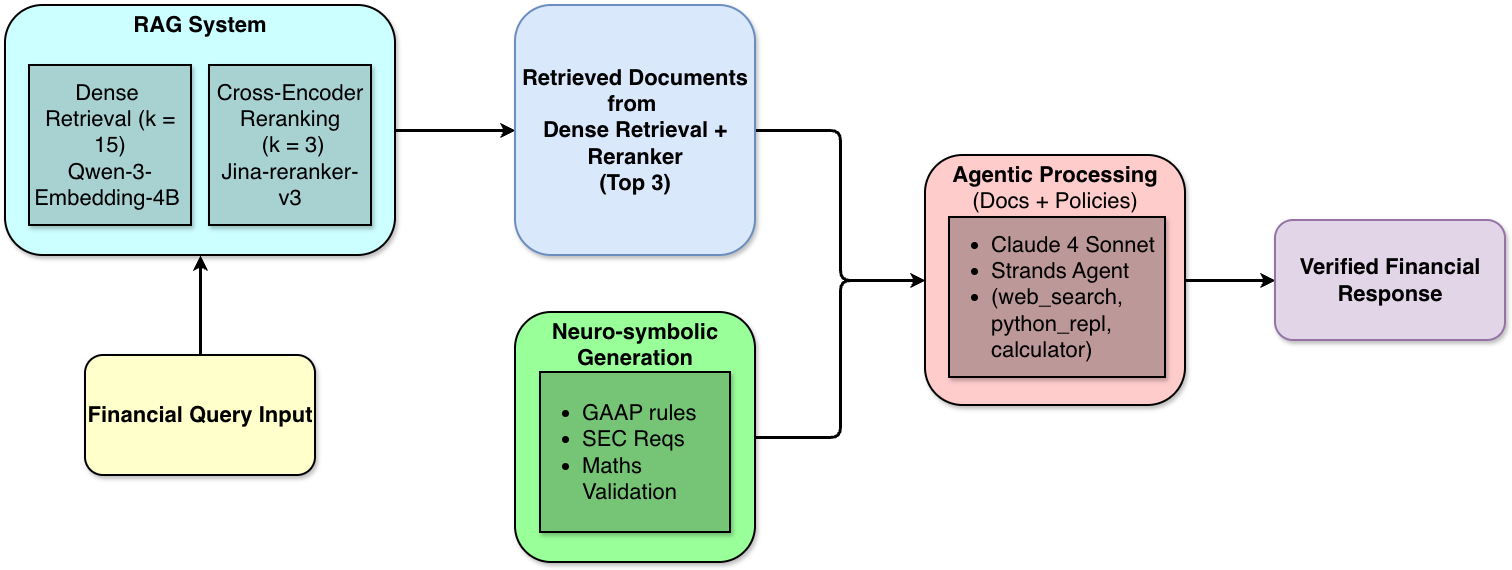}}
\caption{VERAFI system architecture showing the end-to-end pipeline from financial query input through enhanced retrieval (Qwen3-Embedding-4B + Jina-reranker-v3), agentic processing with financial tools, neurosymbolic policy generation using SMT-lib formal specifications for GAAP/SEC constraints, which are embedded as contextual guidelines in the agent's reasoning process to ensure mathematical accuracy and regulatory compliance during generation}
\label{fig:verafi_architecture}
\end{figure*}

\begin{itemize}
\item \textbf{Neurosymbolic Policy Specification:} Application of neurosymbolic autoformalization~\citep{bayless2025neurosymbolic} to translate financial validation requirements into formal SMT-lib specifications, automatically generating rule-based constraints covering GAAP compliance, SEC requirements, and mathematical validation that guide agentic reasoning.
\item \textbf{Policy-Guided Agentic Framework:} Integration of formally-specified policies into agent prompts as contextual guidance, enabling verified reasoning without post-generation validation, ensuring mathematical accuracy and regulatory compliance during generation.
\item \textbf{Empirical Evaluation:} Comprehensive evaluation on FinanceBench-style tasks demonstrating 94.7\% factual correctness (81\% relative improvement over RAG baselines), with neurosymbolic policies contributing 4.3 percentage point gains over pure agentic processing.
\end{itemize}

\section{Related Works}

\subsection{RAG Retrieval Strategies }
The foundation of modern retrieval-augmented generation was established by \citet{lewis2020retrieval}, who introduced the seminal RAG framework that combines pre-trained parametric memory (seq2seq models) with non-parametric memory (dense vector indices of Wikipedia). This work built upon earlier developments in dense passage retrieval, particularly the Dense Passage Retrieval (DPR) method by \citet{karpukhin2020dense}, which demonstrated that learned dense representations could substantially outperform traditional sparse retrieval methods like BM25 by 9-19\% in top-20 passage retrieval accuracy. The RAG approach treats document retrieval as a latent variable problem, marginalizing over retrieved passages to generate more factual and grounded responses. Prior foundational work such as ORQA \citep{lee2019latent} and REALM \citep{guu2020retrieval} had explored similar concepts of integrating retrieval with masked language models, but RAG's generalization to sequence-to-sequence generation tasks marked a significant advancement in making retrieval-augmented approaches applicable to a broader range of knowledge-intensive NLP tasks.

Building upon these foundations, subsequent research has focused on enhancing retrieval quality through advanced reranking strategies and hybrid approaches. \citet{nogueira2019passage} demonstrated the effectiveness of BERT-based cross-encoder reranking, achieving state-of-the-art results on passage ranking benchmarks by fine-tuning BERT to score query-passage relevance more accurately than initial dense retrieval alone. This two-stage approach—fast dense retrieval followed by more computationally expensive but precise neural reranking—has become a standard paradigm in modern RAG systems. Fusion-in-Decoder (FiD) by \citet{izacard2021leveraging} further advanced the field by showing how multiple retrieved passages could be effectively integrated at the decoder level, while large-scale models like RETRO \citep{borgeaud2022improving} demonstrated that retrieval augmentation could enable smaller models to match the performance of much larger parametric-only models. These developments established retrieval quality as a critical bottleneck in RAG systems, with dense retrieval, reranking, and sophisticated fusion mechanisms emerging as essential components for achieving high factual accuracy in knowledge-intensive generation tasks.

\subsection{Advanced and Agentic RAG}

Advanced RAG strategies have evolved beyond basic retrieval-generation pipelines to incorporate sophisticated reasoning, planning, and self-correction mechanisms. A pivotal development was Self-RAG by \citet{asaiself}, which introduced adaptive retrieval through learned reflection tokens that enable models to dynamically decide when to retrieve information and critique their own outputs, significantly improving factual accuracy across knowledge-intensive tasks. Web-based retrieval approaches like WebGPT \citep{nakano2021webgpt} demonstrated how RAG systems could leverage real-time web search rather than static document corpora, enabling access to current information while maintaining generation quality through human feedback training. Active retrieval strategies have emerged through works like Active RAG \citep{jiang2023active}, which optimizes the retrieval process by learning when and what to retrieve based on generation confidence and query complexity. These approaches represent a shift from passive retrieval-augmented generation toward intelligent, self-directed systems capable of strategic information gathering.

Multi-hop reasoning has become a critical frontier in advanced RAG research, addressing queries that require retrieving and reasoning over multiple pieces of supporting evidence. \citet{tang2024multihop} introduced MultiHop-RAG as a comprehensive benchmark revealing that existing RAG methods perform unsatisfactorily on complex queries requiring evidence synthesis from multiple sources, achieving only moderate retrieval accuracy even with reranking techniques. To address these limitations, \citet{trivedi2023interleaving} proposed IRCoT (Interleaved Retrieval with Chain-of-Thought), which alternates between generating reasoning steps and retrieving additional evidence based on partial conclusions, substantially improving both retrieval effectiveness and downstream QA performance. More recently, agentic RAG frameworks \citep{singh2025agentic} have integrated autonomous planning capabilities, tool use, and multi-step reasoning chains, enabling RAG systems to decompose complex queries, orchestrate multiple retrieval operations, and combine diverse information sources through sophisticated planning and execution strategies.

\subsection{Financial AI and Domain-Specific Validation}
Financial AI applications present unique challenges that require specialized validation frameworks due to the high-stakes nature of financial decision-making, regulatory compliance requirements, and the complexity of financial documents. Foundational benchmarks like FinQA \citep{chen2021finqa} and ConvFinQA \citep{chen2022convfinqa} established the importance of numerical reasoning and conversational capabilities in financial question-answering, while FinanceBench \citep{islam2023financebench} provided comprehensive evaluation frameworks that revealed significant limitations in existing AI systems. Recent work has demonstrated these limitations persist, with \citet{wang2025financial} showing that traditional RAG approaches struggle with the exponential growth and complexity of financial data, achieving only modest improvements until domain-specific optimizations are applied. \citet{srinivasan2025enhancing} advanced this field by introducing Multi-HyDE and agentic AI frameworks specifically designed for financial RAG, demonstrating 11.2\% accuracy improvements and 15\% hallucination reduction on financial benchmarks, while highlighting the critical importance of handling intricate regulatory filings and multi-year reports that require sophisticated retrieval strategies beyond standard semantic similarity. The persistent challenges in financial AI validation are further emphasized by \citet{bigeard2025finance}, whose Finance Agent Benchmark revealed that even the most advanced models like OpenAI's o3 achieve only 46.8\% accuracy on real-world financial analysis tasks, underscoring the significant gap between general AI capabilities and the precision required for financial applications that demand specialized validation approaches accounting for numerical precision, temporal relationships, regulatory compliance, and catastrophic risk mitigation.

\subsection{Neurosymbolic AI and Formal Verification}

Neurosymbolic AI combines neural networks with symbolic reasoning to improve interpretability and logical accuracy. Recent work has shown the effectiveness of this approach across diverse domains. \citet{hakim2025symrag} introduced SymRAG, which dynamically selects between symbolic, neural, or hybrid processing based on query complexity. \citet{schmidt2024scaling} applied neurosymbolic approaches to scientific knowledge discovery, showing that Knowledge Graph-based methods outperform purely neural approaches. \citet{allen2025sound} presented a theoretical framework for neurosymbolic reasoning that preserves soundness through paraconsistent logic, while \citet{subramanian2024neuro} demonstrated neurosymbolic approaches for interpretable decision-making in multi-agent scenarios. \citet{bagherinezhad2025enhancing} introduced NeuroSymbolic Augmented Reasoning (NSAR), which extracts symbolic facts from text and generates executable code for reasoning steps.

Recent work on neurosymbolic autoformalization~\citep{bayless2025neurosymbolic} demonstrates how LLMs can translate natural language specifications into formal representations in SMT-lib format~\citep{barrett2010smt}, enabling automated generation of verifiable policies from natural language descriptions. This approach combines neural language understanding with symbolic reasoning to produce machine-checkable specifications through a policy model creator that generates structured rules containing both formal expressions and natural language alternates. Amazon Bedrock Automated Reasoning checks~\citep{akinfaderin2025minimize} implement this autoformalization approach for practical deployment. VERAFI leverages these methods by using autoformalization to generate financial validation policies from domain requirements and incorporating both the formal SMT-lib expressions and natural language alternates into the agent's prompt as contextual guidance during reasoning, rather than applying policies as post-generation validation.
\section{Methodology}

We introduce VERAFI, an agentic framework with neurosymbolic policy generation for financial question-answering that improves accuracy through policy-guided reasoning. Our approach combines two key components: (1) neurosymbolic methods to formally 
specify financial policies in SMT-lib format capturing GAAP and SEC requirements, and (2) incorporating these policies into the agent's prompt as contextual guidance. This enables the system to incorporate domain constraints directly during reasoning.

VERAFI consists of three core components:

\begin{itemize}
\item Dense Retrieval with Reranking
\item Agentic Financial Reasoning
\item Policy-Guided Generation
\end{itemize}

This integrated architecture enables mathematical precision in financial AI applications while maintaining the flexibility and comprehensive coverage needed for diverse financial question-answering scenarios.

\subsection{Dense Retrieval with Reranking}

Our retrieval pipeline employs a two-stage approach that balances recall and precision for financial document retrieval. The first stage performs dense semantic search using Qwen3-Embedding-4B \citep{zhang2025qwen3}, which creates high-dimensional vector representations of both the input query and financial documents in the corpus. The query embedding is generated using a task-specific instruction prompt that guides the model to focus on financial document retrieval, while document embeddings capture the semantic content of financial passages including numerical data, regulatory language, and business context. This initial retrieval step casts a wide net, retrieving the top k=15 documents based on cosine similarity in the embedding space to maximize recall of potentially relevant financial information.

The second stage applies cross-encoder reranking using Jina-reranker-v3 \citep{wang2025jina} to refine the retrieved set and optimize for precision. Unlike the independent encoding used in dense retrieval, the cross-encoder jointly processes the query and each candidate document, enabling more sophisticated relevance modeling that considers query-document interactions. This reranking step reduces the candidate set from 15 documents to the final 3 most relevant passages, significantly improving retrieval quality for complex financial queries that require precise contextual matching. The combination of dense retrieval for broad coverage followed by neural reranking for precision optimization has proven effective for financial question-answering tasks where both comprehensive coverage and accurate relevance assessment are critical.

\subsection{Agentic Financial Reasoning}

  The agentic reasoning could potentially address a fundamental limitation in financial question-answering: while retrieval successfully identifies relevant documents, language models often fail during the reasoning phase when performing complex calculations or in deep domain understanding. Financial queries require table and document understanding, multi-step computations involving ratio calculations, percentage changes across fiscal periods, and compound metrics. Traditional approaches that depend solely on generic LLMs suffer from understanding and arithmetic errors, particularly problematic in financial contexts where precision is essential for regulatory compliance and investment decisions.

  Our baseline system uses the Strands agentic framework\footnote{https://strandsagents.com/latest/} with Claude Sonnet 4 as the base reasoning model, equipped with three computational tools: a symbolic calculator for basic arithmetic operations, a Python REPL environment for complex financial computations, and Tavily\footnote{https://www.tavily.com/} for web search when local document retrieval is insufficient. The agent receives the original financial query $q$ and the top-$k=3$ reranked documents ${d_1, d_2, d_3}$ from the retrieval layer. The agent operates through iterative planning cycles, determining when to extract numerical data from retrieved documents, when to invoke computational tools, and how to structure intermediate reasoning steps. For example, when computing Return on Assets (ROA), the agent extracts net income and
  total assets from the retrieved 10-K documents, invokes the Python REPL to compute the ratio with floating-point precision, and integrates
  the calculated result into its response generation.

  When retrieved documents lack sufficient information to answer a query, the agent automatically invokes Tavily to search for additional financial data from authoritative web sources, ensuring comprehensive coverage beyond the local document corpus.

  This tool-based approach provides several advantages over direct generation. First, arithmetic operations execute with exact precision
  rather than approximate neural predictions. Second, computational steps become transparent and auditable—each tool invocation produces
  explicit intermediate outputs that can be verified by human analysts or downstream validation systems. Third, complex nested calculations
  that would challenge direct language model generation become tractable through sequential tool orchestration. Consider a query requiring
  quick ratio analysis across multiple fiscal years: the agent identifies relevant years from the query, retrieves balance sheet data for
  each period from the reranked documents, computes current assets minus inventory for each year using the Python REPL, divides by current
  liabilities to obtain yearly quick ratios, and synthesizes the temporal trend. Each computational step executes through tool invocation,
  ensuring mathematical accuracy in intermediate results that feed into later reasoning steps.

  The agent is prompted to maintain transparency by explicitly showing computational steps and citing source documents for extracted
  numerical values (see Appendix for full system prompt). The response must include  explicit source citations linking computed values to specific pages and sections within retrieved financial documents, meeting auditability requirements for regulatory compliance. Next, we explore the integration of agentic reasoning with verified computational tools.

\subsection{Policy-Guided Generation}

Our policy-guided generation approach leverages neurosymbolic autoformalization~\citep{bayless2025neurosymbolic} to translate financial domain constraints into formal SMT-lib specifications~\citep{barrett2010smt}, capturing GAAP accounting rules, SEC requirements, and mathematical validation criteria as machine-readable logical specifications. The autoformalization process uses LLM-based translation to convert natural language policy descriptions into structured representations, implemented through Amazon Bedrock Automated Reasoning checks~\citep{akinfaderin2025minimize}. This generates policies containing rules and variables expressed in both SMT-lib format and natural language alternates, covering complex financial relationships such as revenue recognition rules, balance sheet constraints, and regulatory compliance requirements. Rather than applying these policies as post-generation formal verification, we embed both the formal SMT-lib expressions and their natural language alternates directly into the agent's system prompt as contextual guidance.

This approach allows the agent to incorporate formally-specified domain constraints through in-context learning during the reasoning process, guiding it toward mathematical accuracy and regulatory compliance without requiring separate validation layers. The policies, while formally specified using neurosymbolic autoformalization, influence generation through soft constraints embedded in context rather than hard logical constraints applied post-hoc. During generation, the policies are loaded as JSON-formatted rules containing both SMT-lib formal expressions and natural language alternates, providing dual-representation contextual guidance to the Claude Sonnet 4 agent. The agent receives both constraint representations as part of its system context, allowing it to internally validate calculations and conclusions against established financial standards while generating responses. This design choice balances formal rigor in constraint specification with practical efficiency in agentic reasoning, enabling the system to incorporate domain expertise and regulatory knowledge directly into the reasoning process without requiring external validation systems.

\begin{algorithm}[!h]
\caption{VERAFI: Verified Agentic Financial Intelligence}
\begin{algorithmic}[1]
\REQUIRE Financial query $q$, Document corpus $D$, Policy file $P$, Tools $T = \{calculator, python\_repl, tavily\}$, Agent $A$

\STATE \textbf{Policy Loading}
\STATE \COMMENT{Load formally-specified SMT-lib policies}
\STATE Load $policies \leftarrow$ LoadJSON($P$)
\STATE $rules \leftarrow policies.rules[0:max\_rules]$ \COMMENT{Select subset for context limits}
\STATE $P_{context} \leftarrow$ ``FINANCIAL VALIDATION RULES:''
\STATE \COMMENT{Format policy rules for context inclusion}
\FOR{$rule$ in $rules$}
    \STATE $P_{context} \leftarrow P_{context} + rule.alternateExpression$
\ENDFOR
\STATE 
\STATE \textbf{Stage 1: Enhanced Retrieval}
\STATE $query_{embed} \leftarrow$ Qwen3Embedding($q$) \COMMENT{Task-specific embedding}
\STATE $docs_{dense} \leftarrow$ DenseSearch($query_{embed}$, $D$, $k=15$) \COMMENT{Semantic similarity}
\STATE $doc\_texts \leftarrow$ ExtractContent($docs_{dense}$)
\STATE $docs_{final} \leftarrow$ JinaReranker($q$, $doc\_texts$, $k=3$) \COMMENT{Cross-encoder reranking}
\STATE
\STATE \textbf{Stage 2: Policy-Guided Generation}
\STATE $instructions \leftarrow$ ``Use policy guidelines to verify calculations. Cite sources clearly.''
\STATE $prompt \leftarrow P_{context} + q + context + instructions$
\STATE $response \leftarrow$ Agent($prompt$, $T$) \COMMENT{Claude Sonnet 4 with financial tools}
\STATE
\STATE \textbf{Stage 3: Answer Extraction}
\IF{$response$ has $message$ attribute}
    \STATE $answer \leftarrow response.message$
\ELSIF{$response$ is dictionary}
    \FOR{$key$ in $\{message, content, text, answer\}$}
        \IF{$key$ in $response$}
            \STATE $answer \leftarrow response[key]$
            \STATE break
        \ENDIF
    \ENDFOR
\ELSE
    \STATE $answer \leftarrow$ string($response$)
\ENDIF
\STATE
\RETURN $answer$ \COMMENT{Policy-guided financial response}

\end{algorithmic}
\end{algorithm}

\section{Experimental Setup}

We evaluate VERAFI on financial question-answering tasks using FinanceBench-style datasets to assess both retrieval effectiveness and generation quality. Our evaluation framework examines retrieval performance through standard information retrieval metrics and generation quality through factual correctness and completeness assessments across multiple system configurations.

\subsection{Evaluation Dataset and Metrics}

\textbf{Dataset Construction:} Following the methodology established by \citet{srinivasan2025enhancing}, we construct our evaluation dataset using a targeted subset of companies from FinanceBench \citep{islam2023financebench} and ConvFinQA \citep{chen2022convfinqa}. Our dataset focuses on four major corporations: American Water Works (2015-2018 filings), AMD (2022 filings), American Express (2022 filings), and Boeing (2022 filings). This selection provides diverse financial contexts including utilities, technology, financial services, and aerospace industries, enabling comprehensive evaluation across different financial domains and reporting periods.

The document corpus consists of SEC filings (10-K, 10-Q, 8-K reports) processed into a vector database using Qwen3-Embedding-4B embeddings. Financial documents are segmented using recursive character splitting with 500-character chunks and 50-character overlap to preserve contextual information while maintaining retrieval granularity. The final corpus contains financial passages with metadata including document names, page numbers, and source references, stored in a Chroma vector database for efficient similarity search.

\textbf{Evaluation Metrics:} We assess VERAFI performance using two complementary evaluation frameworks. For retrieval effectiveness, we employ 
standard information retrieval metrics including Recall@3, NDCG@3, MRR@3, and Hit Rate@3, measuring the system's ability to identify relevant financial passages for complex queries. For generation quality assessment, we utilize LLM-as-a-Judge evaluation \citep{liu2023g, zheng2023judging} to measure factual correctness and completeness of generated financial responses. We implement this evaluation methodology using Amazon Bedrock's LLM-as-a-Judge framework \citep{akinfaderin2025llmasjudge}, which provides pre-trained evaluator models with optimized prompt engineering. The judge model evaluates whether responses accurately reflect the retrieved financial information and provide answers to the posed questions, enabling systematic assessment of the generation components within the integrated VERAFI framework.

\subsection{Experimental Results}
We present evaluation results examining both retrieval effectiveness and generation quality across multiple system configurations, demonstrating the improvements achieved through better retrieval strategies and agentic processing.

\subsubsection{Retrieval Performance}

Table \ref{tab:retrieval_results} presents retrieval performance across six different methods evaluated on our financial QA dataset. The results show clear performance differences and reveal important insights about retrieval strategies for financial documents.

\begin{table*}
\caption{Retrieval performance comparison across different methods on financial QA dataset. Results show Recall@3, NDCG@3, MRR@3, and Hit Rate@3 metrics. Best performance for each metric is highlighted in bold. Dense+Rerank consistently outperforms all other approaches.}
\label{tab:retrieval_results}
\centering
\begin{tabular}{@{}lcccc@{}}
\hline
\textbf{Method} & \textbf{Recall@3} & \textbf{NDCG@3} & \textbf{MRR@3} & \textbf{Hit@3} \\
\hline
Dense+Rerank & \textbf{0.667} & \textbf{0.486} & \textbf{0.429} & \textbf{0.619} \\
Dense & 0.381 & 0.277 & 0.238 & 0.381 \\
BM25+Rerank & 0.190 & 0.101 & 0.095 & 0.143 \\
Hybrid+Rerank & 0.190 & 0.113 & 0.119 & 0.143 \\
BM25 & 0.095 & 0.038 & 0.024 & 0.048 \\
Hybrid & 0.095 & 0.075 & 0.063 & 0.095 \\
\hline
\end{tabular}
\end{table*}

\textbf{Key Findings:} Dense+Rerank achieves best performance, dominating across all metrics with 66.7\% recall, successfully retrieving relevant documents for approximately two-thirds of financial queries. Cross-encoder reranking provides substantial improvements, particularly for dense retrieval (0.381 → 0.667 recall improvement). BM25 shows limited effectiveness on financial texts, likely due to technical terminology and numeric data complexity, while hybrid methods underperform simple dense retrieval, indicating the need for better fusion approaches.

\textbf{Implications for Financial RAG Systems:} Even optimized retrieval methods achieve only 66.7\% recall, leaving room for improvement through validation and correction mechanisms. The 33.3\% of queries with poor retrieval represent critical cases where post-generation validation becomes essential, as LLMs must work with incomplete or incorrect context. Dense+Rerank represents the current baseline for subsequent agentic and neurosymbolic improvements.

\subsubsection{Generation Quality}

Table \ref{tab:generation_results} presents generation quality results across different system configurations, evaluated using LLM-as-a-Judge assessment for factual correctness and completeness. The results demonstrate significant improvements from our agentic RAG approach compared to baseline retrieval methods.

\begin{table*}
\caption{Generation quality comparison across system configurations on financial QA dataset. Factual correctness and completeness are evaluated using LLM-as-a-Judge methodology. Best performance for each metric is highlighted in bold.}
\label{tab:generation_results}
\centering
\begin{tabular}{@{}lcc@{}}
\hline
\textbf{Method} & \textbf{Factual Correctness} & \textbf{Completeness} \\
\hline
RAG (Dense) + Reranker + Agent (+ Web Search) + Neurosymbolic & \textbf{0.947} & \textbf{0.964} \\
RAG (Dense) + Reranker + Agent (+ Web Search) & 0.904 & \textbf{0.964} \\
Dense + Reranker & 0.524 & 0.726 \\
Only Dense & 0.333 & 0.536 \\
Hybrid + Reranker & 0.333 & 0.500 \\
BM25 + Reranker & 0.190 & 0.500 \\
Only BM25 & 0.071 & 0.440 \\
Only Hybrid & 0.048 & 0.583 \\
\hline
\end{tabular}
\end{table*}

\begin{figure}[h]
  \centering
  \includegraphics[width=\columnwidth]{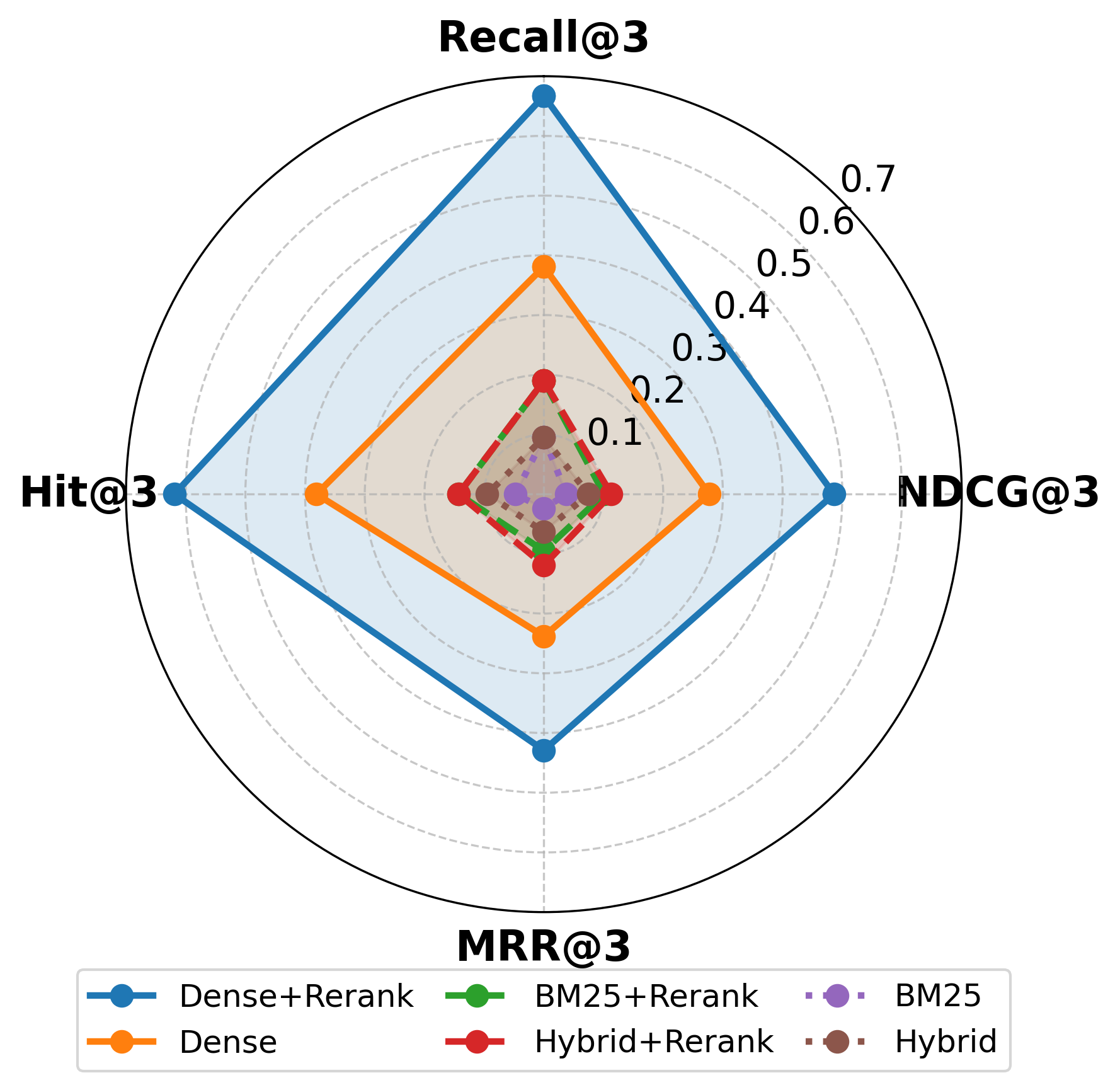}
  \caption{Retrieval performance comparison across different methods at k=3. Dense+Rerank substantially outperforms all other approaches
  across all metrics (Recall@3, NDCG@3, MRR@3, Hit@3), demonstrating the effectiveness of combining dense retrieval with cross-encoder
  reranking for financial document retrieval.}
  \label{fig:retrieval_radar}
  \end{figure}

Our best-performing configuration, VERAFI with neurosymbolic validation, achieves 94.7\% factual correctness and 96.4\% completeness, substantially outperforming all baseline methods. This represents a dramatic improvement over traditional approaches, with factual correctness increasing from 52.4\% (Dense + Reranker) to 90.4\% with agentic processing and web search, and reaching 94.7\% with neurosymbolic validation. Notably, these results significantly exceed the performance reported by \citet{srinivasan2025enhancing} on similar financial QA tasks, where the highest factual correctness achieved was 38.49\%\footnote{Evaluation methodologies differ: \citet{srinivasan2025enhancing} used RAGAS framework with GPT-4, while our evaluation employs Amazon Bedrock Evals LLM-as-a-Judge with Claude 3.7 Sonnet v1, which may contribute to performance differences.}.

The neurosymbolic policy layer contributes a meaningful 4.3 percentage point improvement beyond the agentic baseline, demonstrating the effectiveness of incorporating financial domain expertise directly into the reasoning process to target mathematical and logical errors that persist even after retrieval optimization.
\begin{figure}[!h]
  \centering
  \includegraphics[width=\columnwidth]{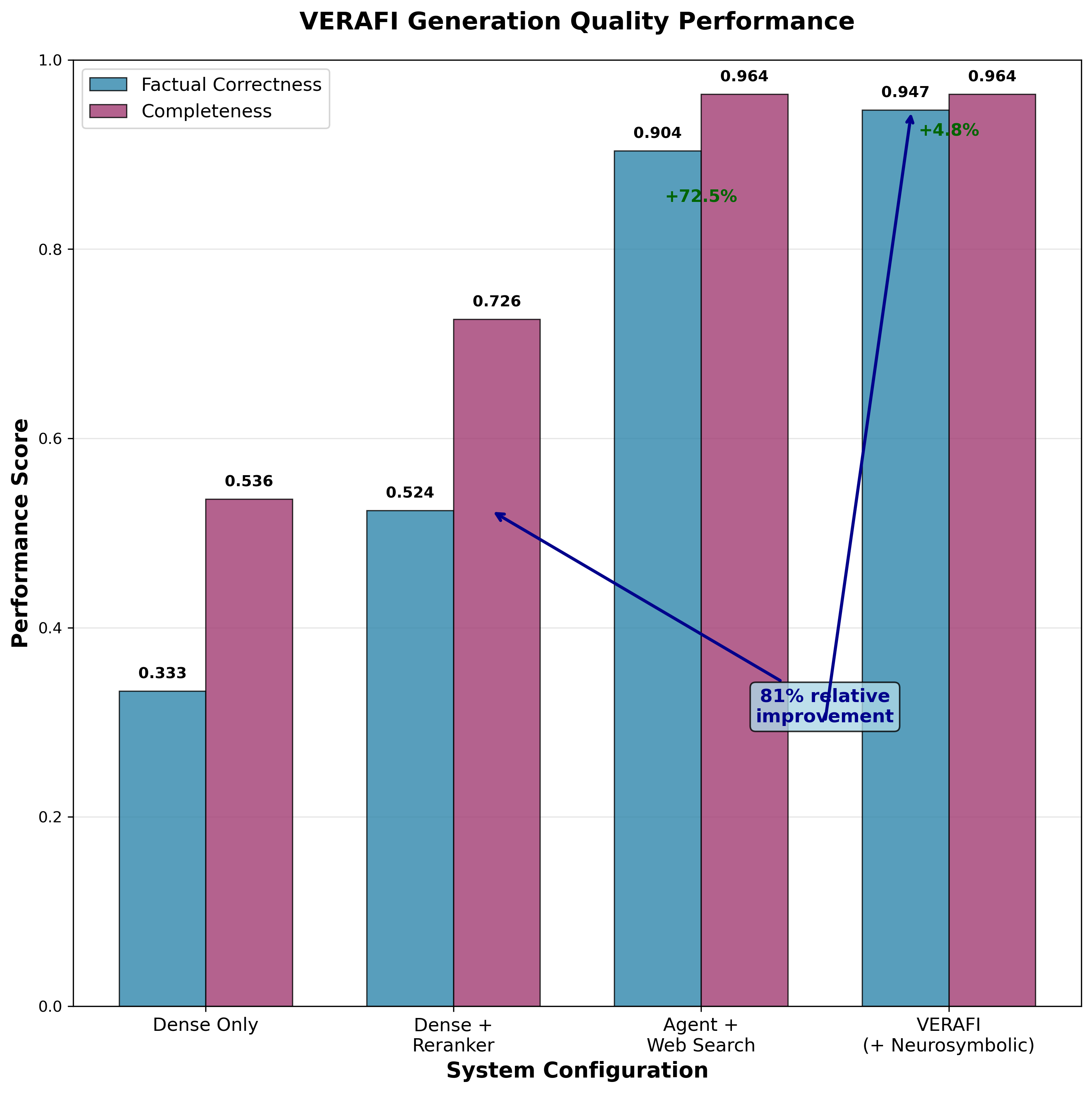}
  \caption{Generation quality performance across VERAFI system configurations. Factual correctness improves from 52.4\% (Dense + Reranker) to 94.7\% (full VERAFI), with agentic processing providing the largest gain and neurosymbolic validation contributing an additional 4.3 percentage points.}
  \label{fig:generation_quality}
\end{figure}

\section{Conclusion}

This paper introduces VERAFI, a neurosymbolic framework that combines enhanced retrieval-augmented generation with agentic and neurosymbolic framework for verified financial intelligence. Through evaluation on FinanceBench-style datasets (ConvFinQA), we demonstrated that integrating agentic processing with automated reasoning policies can significantly improve accuracy in financial question-answering systems. Our results show that VERAFI achieves 94.7\% factual correctness, representing an 81\% relative improvement over traditional dense retrieval with reranking approaches. The neurosymbolic validation layer contributes meaningful improvements beyond pure agentic processing, demonstrating that including financial domain expertise directly into the reasoning process effectively targets mathematical and logical errors that persist even after retrieval optimization.

Our work offers immediate practical value for financial institutions working with high-stakes AI applications where accuracy is paramount for regulatory compliance and investment decisions. The VERAFI framework provides a deployable solution that addresses the critical gap between general-purpose RAG systems and the precision requirements of financial AI through computational tools, web search capabilities, and domain-specific validation policies. Future research directions include expanding the automated reasoning policy framework to additional financial domains, investigating dynamic policy selection based on query complexity, and exploring the integration of VERAFI with real-time market data streams for enhanced financial decision support systems.

\clearpage
\onecolumn

\section{Appendix}

  \subsection{Baseline Agent System Prompt}

  The baseline agent system (without neurosymbolic validation) uses the following prompt to guide financial analysis:

  \begin{listing}[h]
  \begin{lstlisting}[breaklines=true, basicstyle=\small\ttfamily]
  You are a financial analyst. Answer this question using ONLY the provided documents.

  Question: {question}

  Retrieved Financial Documents:
  {formatted_docs}

  Instructions:
  - Use only the information in the provided documents
  - Use calculator for basic math operations
  - Use python_repl for complex calculations or data analysis
  - Show your calculations clearly
  - Cite document sources in your answer
  - If information is missing, state that clearly

  Provide a complete analysis with calculations and citations.
  \end{lstlisting}
  \end{listing}

  \subsection{Neurosymbolic Agent System Prompt (In-Context)}

  The neurosymbolic agent system incorporates financial validation rules directly into the prompt context:

  \begin{listing}[h]
  \begin{lstlisting}[breaklines=true, basicstyle=\small\ttfamily]
  You are a professional financial analyst. Answer this question using the provided documents.

  FINANCIAL VALIDATION RULES (for internal use only):

  {rule['alternateExpression'] for each policy rule}
  ... and {N} additional validation rules

  Question: {question}

  Retrieved Financial Documents:
  {formatted_docs}

  Instructions:
  - Provide a clear, concise financial analysis based solely on the provided documents
  - Use calculator for basic math operations and python_repl for complex calculations
  - Use the validation rules above to internally verify your calculations (DO NOT mention rule IDs or validation details in your response)
  - Present a professional analysis focused on business insights
  - Cite document sources clearly
  - If calculations don't align with validation rules, note any discrepancies briefly
  - Keep your response focused and avoid unnecessary technical details

  Your response should read like a professional financial report, not a technical validation log.
  \end{lstlisting}
  \end{listing}

  \subsection{Generation Prompt for RAG-Only Baseline}

  The RAG-only baseline (without agentic tools) uses Claude Sonnet 4 with the following prompt structure:

  \begin{listing}[h]
  \begin{lstlisting}[breaklines=true, basicstyle=\small\ttfamily]
  You are a financial analyst. Answer the following question using ONLY the information provided in the retrieved documents.

  Question: {query}

  Retrieved Documents:
  {context}

  Instructions:
  - Provide specific numerical answers when requested
  - Cite the document source when possible
  - If the information is not available in the documents, state that clearly
  - Show calculations when relevant
  - Focus only on information from the retrieved documents

  Answer:
  \end{lstlisting}
  \end{listing}

\subsection{Financial Validation Policies}

VERAFI embeds formally-specified financial policies into the agent's reasoning 
context. These policies are generated using neurosymbolic methods and stored in 
JSON format, with each rule containing two representations: (1) the formal 
SMT-lib specification \citep{barrett2010smt} for logical rigor, and (2) a 
natural language alternate expression for in-context integration in the agent's 
prompt. 

The following are key financial validation policies covering GAAP accounting 
standards, SEC regulatory requirements, and mathematical validation criteria. 
These examples are representative of the complete 80+ rule set that guides 
the agent's reasoning during generation:

\vspace{-0.2cm}
\begin{listing*}[h]
\begin{lstlisting}[language=json, breaklines=true]
{
    "rules": [
        {"id": "ID8", "alternateExpression": "returnOnAssets is equal to netIncome / averageTotalAssets", "expression": "(= returnOnAssets (/ netIncome averageTotalAssets))"},
        {"id": "ID9", "alternateExpression": "debtToEquityRatio is equal to totalDebt / totalShareholdersEquity", "expression": "(= debtToEquityRatio (/ totalDebt totalShareholdersEquity))"},
        {"id": "ID11", "alternateExpression": "currentRatio is equal to currentAssets / currentLiabilities", "expression": "(= currentRatio (/ currentAssets currentLiabilities))"},
        {"id": "ID15", "alternateExpression": "freeCashFlow is equal to operatingCashFlow - capitalExpenditures", "expression": "(= freeCashFlow (- operatingCashFlow capitalExpenditures))"},
        {"id": "ID19", "alternateExpression": "if dataSource is equal to SEC_FILING, then usesMostAuthoritativeSource is true", "expression": "(=> (= dataSource SEC_FILING) usesMostAuthoritativeSource)"}
    ]
}
\end{lstlisting}
\end{listing*}
\vspace{-0.2cm}
\end{document}